%% file: HALF.tex
\documentclass[a4paper,conference]{ieee/IEEEtran}

\usepackage{amssymb}
\usepackage[cmex10]{amsmath}
\usepackage{amsfonts}
\usepackage{float}
\usepackage[square,sort,comma,numbers]{natbib}
\usepackage{graphicx}
\usepackage{subcaption}
\usepackage{calc}
\usepackage{wrapfig}
\usepackage{tabularx}
\usepackage{booktabs}
\usepackage{footnote}
\usepackage{makeidx}
\usepackage{balance}%
\usepackage{textcomp}
\usepackage{booktabs}
\usepackage{mathtools}
\usepackage{amsmath}
\usepackage{threeparttable}
\usepackage{multirow}
\usepackage{rotating}
\usepackage{color}
\usepackage{makecell}
\usepackage{url}
\usepackage{array}
\usepackage{setspace}
\usepackage{siunitx}
\usepackage[table]{xcolor}
\usepackage{color, colortbl}
\usepackage[capitalize]{cleveref}
\usepackage{etoolbox}
\apptocmd{\thebibliography}{\raggedright}{}{}
\usepackage{caption}
\usepackage{siunitx}
\setcitestyle{square}
\usepackage{comment}
\usepackage[super]{nth}
\usepackage{scalerel}

\usepackage[acronym]{glossaries}
\newglossary[symlog]{symbol}{symi}{symo}{Symbols}
\glstoctrue 
\loadglsentries[\acronymtype]{tex/acronyms}

\renewcommand{\baselinestretch}{0.93}

\begin{document}

\title{HALF: \underline{H}olistic \underline{A}uto Machine \underline{L}earning for \underline{F}PGAs}

\author{%
  \IEEEauthorblockN{%
    Jonas Ney\IEEEauthorrefmark{1}\textsuperscript{\textsection},
    Dominik Loroch\IEEEauthorrefmark{2}\textsuperscript{\textsection},
    Vladimir Rybalkin\IEEEauthorrefmark{1}\textsuperscript{\textsection},
    Nico Weber\IEEEauthorrefmark{2},
    Jens Krüger\IEEEauthorrefmark{2} and
    Norbert Wehn\IEEEauthorrefmark{1}%
  }%
  \IEEEauthorblockA{\IEEEauthorrefmark{1} University of Kaiserslautern, Kaiserslautern, Germany}%
  \IEEEauthorblockA{\IEEEauthorrefmark{2} Fraunhofer ITWM, Kaiserslautern, Germany\\
  \{ney, rybalkin, wehn\}@eit.uni-kl.de, \{dominik.loroch, nico.weber, jens.krueger\}@itwm.fraunhofer.de}%
}

\maketitle

\begingroup\renewcommand\thefootnote{\textsection}
\footnotetext{first three authors contributed equally}
\endgroup

\newcommand\blfootnote[1]{%
    \begingroup
    \renewcommand\thefootnote{}\footnote{#1}%
    \addtocounter{footnote}{-1}%
    \endgroup
}

\blfootnote{DOI: 10.1109/FPL53798.2021.00069}

\input{tex/abstract_neu_Norbert}

\vspace{2ex}

\input{tex/instroduction_neu_Norbert}
\input{tex/related_works}
\input{tex/framework_neu_Norbert}

\input{tex/hardware_aware_objectives}
\input{tex/hardware_library}

\input{tex/results_neu_Norbert}

\input{tex/conclusion_neu_Norbert}

\section*{Acknowledgement}

This work has been funded by the German Federal Ministry of Education and Research as a participant of the pilot innovation competition "Energy-efficient AI System".

\bibliographystyle{ieee/IEEEtran} 
\bibliography{ieee/IEEEexample}

\end{document}

%% file: tex/acronyms.tex
\newcommand{\newacronymf}[3]{\newglossaryentry{#1}{name={#2},description={#3},first={#2}}}

\newacronymf{3gpp}{3GPP}{Third Generation Partnership Project}
\newacronymf{4g}{4G}{fourth generation of mobile phone system specifications}
\newacronymf{5g}{5G}{fifth generation of mobile phone system specifications}
\newacronym{1dlstm}{1D-LSTM}{One-dimensional Long Short-Term Memory}
\newacronym{2dlstm}{2D-LSTM}{2D Long Short-Term Memory}
\newacronym{1dcnn}{1D-CNN}{1D Convolutional Neural Network}
\newacronym{2dcnn}{2D-CNN}{2D Convolutional Neural Network}

\newacronym{ann}{ANN}{Artificial Neural Network}
\newacronym{anns}{ANNs}{Artificial Neural Networks}
\newacronym{acq}{ACQ}{acquisition}
\newacronym{app}{APP}{a posteriori probability}
\newacronym{ask}{ASK}{amplitude-shift keying}
\newacronym{awgn}{AWGN}{Additive White Gaussian Noise}
\newacronym{arq}{ARQ}{automatic repeat request}
\newacronym{acs}{ACS}{add-compare-select}
\newacronym{asic}{ASIC}{application specific integrated circuit}
\newacronym{arp}{ARP}{almost regular permuatation}
\newacronym{asip}{ASIP}{application specific instruction set processor}
\newacronymf{amba}{AMBA}{Advanced Microcontroller Bus Architecture}
\newacronym{ahb}{AHB}{advanced high-performance bus}
\newacronym{aacs}{AACS}{a priori aware constant sphere}
\newacronym{ant}{ANT}{algorithmic noise-tolerance}
\newacronym{apu}{APU}{Application Processing Unit}

\newacronym{bhl}{BHL}{Backward Hidden Layer}
\newacronym{blstm}{BLSTM}{Bidirectional Long Short-Term Memory}
\newacronym{bcjr}{BCJR}{Bahl, Cocke, Jelinek, and Raviv}
\newacronym{bpsk}{BPSK}{Binary Phase Shift Keying}
\newacronym{ber}{BER}{bit error rate}
\newacronym{bp}{BP}{Belief Propagation}
\newacronym{bw}{BW}{Bit Width}
\newacronym{bicm}{BICM}{bit interleaved coded modulation}
\newacronym{blast}{BLAST}{Bell Labs layered space time}
\newacronymf{bc}{BC}{best case, 1.3V, -40$^\circ$C}
\newacronym{bilstm}{Bi-LSTM}{Bidirectional LSTM}

\newacronym{ctc}{CTC}{Connectionist Temporal Classification}
\newacronym{cn}{CN}{Check Node}
\newacronym{cr}{CR}{Code Rate}
\newacronym{cfu}{CFU}{Check Node Functional Unit}
\newacronym{crc}{CRC}{Cyclic Redundancy Check}
\newacronymf{cdma2000}{CDMA2000}{name of a communication standard}
\newacronym{cc}{CC}{convolutional code}
\newacronym{cnb}{CNB}{check node block}
\newacronym{cnp}{CNP}{check node processor}
\newacronym{cdma}{CDMA}{code division multiple access}
\newacronymf{cmos}{CMOS}{complementary metal oxide semiconductor}
\newacronymf{cnn}{CNN}{Convolutional Neural Network}
\newacronymf{cer}{CER}{Character Error Rate}
\newacronym{cpu}{CPU}{Central Processing Unit }
\newacronym{cgp}{CGP}{Cartesian Genetic Programming }

\newacronym{darts}{DARTS}{differentiable architecture search}
\newacronym{dnnu}{DNNU}{Deep Neural Network Unit}
\newacronym{dma}{DMA}{Direct Memory Access}
\newacronym{db}{dB}{decibel}
\newacronym{dp}{DP}{dual ported (memory)}
\newacronym{dram}{DRAM}{dynamic random access memory}
\newacronym{dc}{$d_c$}{CN degree}
\newacronym{dv}{$d_v$}{VN degree}
\newacronym{dvs}{DVS}{dynamic voltage scaling}
\newacronym{dvb}{DVB}{digital video broadcast}
\newacronymf{dvbrcs}{DVB-RCS}{digital video broadcasting - return channel over satellite}
\newacronymf{dvbrct}{DVB-RCT}{digital video broadcasting - return channel terrestrial}
\newacronymf{dvbc2}{DVB-C2}{digital video broadcasting - cable - second generation \cite{eudigital}}
\newacronymf{dvbs2}{DVB-S2}{digital video broadcasting - satellite - second generation \cite{eudigital}}
\newacronymf{dvbs2x}{DVB-S2X}{digital video broadcasting - satellite - extension of the second generation \cite{eudigital}}
\newacronymf{dvbt2}{DVB-T2}{digital video broadcasting - terrestrial - second generation \cite{eudigital}}
\newacronym{dvfs}{DVFS}{dynamic voltage and frequency scaling}
\newacronym{drp}{DRP}{dithered relative prime}
\newacronymf{dfg}{DFG}{Deutsche Forschungsgesellschaft}
\newacronym{dnn}{DNN}{Deep Neural Network}
\newacronym{dart}{DART}{Distributed Analytics Runtime}
\newacronym{dl}{DL}{deep learning}

\newacronym{esf}{ESF}{extrinsic scaling factor}
\newacronym{edc}{EDC}{error detection and correction}
\newacronym{edge}{EDGE}{name of a communications standard}
\newacronym{eds}{EDS}{error detection sequential}
\newacronym{ecc}{ECC}{error correction code}
\newacronym{exit}{EXIT}{extrinsic information transfer}
\newacronym{ems}{EMS}{Extended Min-Sum}
\newacronym{ecn}{ECN}{Elementary Check Node}
\newacronym{ecg}{ECG}{Electrocardiogram}

\newacronym{fhl}{FHL}{Forward Hidden Layer}
\newacronym{fsm}{FSM}{Finite-State Machine}
\newacronym{fec}{FEC}{forward error correction}
\newacronym{fer}{FER}{Frame Error Rate}
\newacronymf{fifo}{FIFO}{first in first out memory}
\newacronym{fsd}{FSD}{Fixed complexity sphere decoder}
\newacronym{fwbw}{FWBW}{Forward-Backward}
\newacronym{fcn}{FCN}{Fully Convolutional Network}
\newacronym{fpn}{FPN}{Feature Pyramid Network}
\newacronym{fpga}{FPGA}{Field-Programmable Gate Array}

\newacronymf{gsm}{GSM}{Global System for Mobile Communication; communication standard}
\newacronym{gpp}{GPP}{general-purpose processor}
\newacronym{gprs}{GPRS}{name of a communications standard}
\newacronym{gps}{GPS}{global positioning system}
\newacronymf{geq}{GE}{gate equivalent}
\newacronym{gfq}{GF$(q)$}{Galois Field}
\newacronym{gru}{GRU}{gated recurrent unit}
\newacronym{gpu}{GPU}{graphics processor unit}

\newacronymf{hspa}{HSPA}{high speed packet access}
\newacronym{half}{HALF}{\underline{H}olistic \underline{A}uto machine \underline{L}earning for \underline{F}PGAs}
\newacronym{harq}{HARQ}{hybrid automatic repeat request}
\newacronym{hw}{HW}{hardware}
\newacronym{hls}{HLS}{High-level Synthesis}
\newacronym{hpc}{HPC}{high performance computing}

\newacronym{ic}{IC}{integrated circuit}
\newacronym{io}{I/O}{input / output}
\newacronym{ip}{IP}{Intellectual Property}
\newacronym{ier}{IER}{injected error rate}
\newacronym{itrs}{ITRS}{International Technology Roadmap for Semiconductors}


\newacronym{knn}{KNNs}{künstliche neuronale Netze}

\newacronym{lstm}{LSTM}{Long Short-Term Memory}
\newacronym{llr}{LLR}{Logarithmic Likelihood Ratio}
\newacronym{ldr}{LDR}{Logarithmic Density Ratio}
\newacronymf{lte}{LTE}{long term evolution of the 3GPP standard}
\newacronym{ldpc}{LDPC}{Low-Density Parity-Check}
\newacronymf{logmap}{Log-MAP}{MAP algorithm in the logarithmic domain}
\newacronym{lan}{LAN}{local area network }
\newacronym{lsb}{LSB}{least significant bit}
\newacronym{lut}{LUT}{look-up table}
\newacronym{lfsr}{LFSR}{linear feedback shift register}
\newacronym{lord}{LORD}{layered orthogonal lattice detector}
\newacronym{lsd}{LSD}{List sphere decoder}
\newacronym{ldmc}{LDMC}{low-density MIMO check}

\newacronym{mac}{MAC}{Multiply-Accumulate}
\newacronym{map}{MAP}{maximum a posteriori probability}
\newacronymf{mlm}{Max-Log-MAP}{MAP algorithm in the logarithmic domain with simplified computation}
\newacronym{mimo}{MIMO}{multiple input multiple output}
\newacronym{mtbf}{MTBF}{mean time between failure}
\newacronym{msb}{MSB}{most significant bit}
\newglossaryentry{mpsoc}{name=MPSoC, description=multi-processor system-on-chip, first=multi-processor system-on-chip (MPSoC), firstplural=multi-processor systems-on-chip (MPSoC)}
\newacronym{mops}{MOPS}{million operations per second}
\newacronym{mmse}{MMSE}{minimum mean square error}
\newacronym{mcmc}{MCMC}{Markov Chain Monte Carlo}
\newacronym{mlh}{ML}{maximum likelihood}
\newacronym{ml}{ML}{Machine Learning}
\newacronym{mdlstm}{MD-LSTM}{Multidimensional Long Short-Term Memory}
\newacronym{mlp}{MLP}{Multilayer Perceptron}

\newacronym{nas}{NAS}{Neural Architecture Search}
\newacronym{nns}{NNs}{Neural Networks}
\newacronym{nn}{NN}{Neural Network}
\newacronym{nbldpc}{NB-LDPC}{Non-Binary Low-Density Parity-Check}
\newacronym{nii}{NII}{next iteration initialization}
\newacronym{nsc}{NSC}{non-systematic and non-recursive convolutional}
\newacronym{noc}{NoC}{network on chip}
\newacronymf{nom}{NOM}{nominal case, 1.2V, 25$^\circ$C}

\newacronym{ocr}{OCR}{Optical Character Recognition}
\newacronym{otx}{OTX}{outer receiver}
\newacronym{ofdm}{OFDM}{orthogonal frequency division multiplexing}

\newacronym{pl}{PL}{Programmable Logic}
\newacronym{ps}{PS}{Processing System}
\newacronym{psk}{PSK}{phase-shift keying}
\newacronym{psmap}{PSMAP}{parallel SMAP}
\newacronym{pr}{P\&R}{placement and routing}
\newacronym{ped}{PED}{partial Euclidean distance}
\newacronym{pic}{PIC}{parallel interference cancellation}
\newacronym{pn}{PN}{Permutation Node}

\newacronym{qam}{QAM}{quadrature amplitude modulation}
\newacronym{qpsk}{QPSK}{quadrature phase shift keying}
\newacronym{qpp}{QPP}{quadratic permutation polynomial}
\newacronym{qos}{QoS}{Quality of Service}

\newacronym{rl}{RL}{reinforcement learning}
\newacronym{rnns}{RNNs}{Recurrent Neural Networks}
\newacronym{rnn}{RNN}{Recurrent Neural Network}
\newacronym{rsc}{RSC}{recursive systematic convolutional}
\newglossaryentry{ram}{name=RAM,description=random access memory,first=random access memory (RAM),firstplural=random access memories (RAMs)}
\newglossaryentry{rom}{name=ROM,description=read only memory,first=read only memory (ROM),firstplural=read only memories (ROMs)}
\newacronym{rms}{RMS}{recognition, mining, and synthesis}
\newacronym{rs}{RS}{reed-solomon code}
\newacronym{rtl}{RTL}{register transfer level}
\newacronym{rts}{RTS}{repeated tree search}
\newacronym{rap}{RAP}{Resilience Articulation Point}

\newacronym{snr}{SNR}{Signal-to-Noise Ratio}
\newglossaryentry{sram}{name=SRAM,description=static random access memory,first=static random access memory (SRAM),firstplural=static random access memories (SRAMs)}
\newacronym{smap}{SMAP}{serial MAP}
\newacronym{sdr}{SDR}{software-defined radio}
\newacronymf{sdram}{SDRAM}{synchronous dynamic random access memory}
\newacronym{set}{SET}{single-event transient}
\newacronym{seu}{SEU}{single-event upset}
\newacronym{siso}{SISO}{soft-input soft-output}
\newacronym{siso2}{SISO}{single-input single-output}
\newacronym{soc}{SoC}{Systems-on-Chip}
\newacronym{sova}{SOVA}{soft-output Viterbi algorithm}
\newacronym{sw}{SW}{software}
\newacronym{sm}{SM}{spatial multiplexing}
\newacronym{stbc}{STBC}{space-time block code}
\newacronym{sgd}{SGD}{stochastic gradient descend}
\newacronym{sttc}{STTC}{space-time trellis code}
\newacronym{sic}{SIC}{successive interference cancellation}
\newacronym{sts}{STS}{single tree search}
\newacronymf{spp}{SPP}{Schwerpunktprogramm}
\newacronym{se}{SE}{Schnorr-Euchner}
\newacronym{synb}{SYN}{Syndrome-Based}

\newacronym{tdp}{TDP}{thermal design power}
\newacronym{tmr}{TMR}{triple modular redundancy}
\newacronym{tc}{TC}{turbo code}
\newacronym{ts}{TS}{Tuple search}

\newacronymf{umts}{UMTS}{universal mobile telecommunications system}
\newacronymf{uwb}{UWB}{ultra-wide band}
\newacronym{usb}{USB}{universal serial bus}
\newacronym{umic}{UMIC}{ultra high speed mobile information and communication}

\newacronym{vn}{VN}{Variable Node}
\newacronym{vfu}{VFU}{variable node functional unit}
\newacronym{vhdl}{VHDL}{very high speed integrated circuits hardware description language}
\newacronym{vliw}{VLIW}{very large instruction word}
\newacronym{vnb}{VNB}{variable node block (in \gls{LDPC} decoder architectures)}
\newacronym{vnp}{VNP}{variable node processors (in \gls{LDPC} decoder architectures)}
\newacronym{vblast}{V-BLAST}{Vertical Bell Labs layered space time}

\newacronymf{wimax}{WiMAX}{worldwide interoperability for microwave access \cite{ieworldwide}}
\newacronymf{wimedia}{WiMedia}{name of an alliance for standardization}
\newacronymf{wifi}{WiFi}{trademark of Wi-Fi alliance}
\newacronymf{wc}{WC}{worst case, 1.1V, 125$^\circ$C}
\newacronymf{wer}{WER}{Word Error Rate}

\newglossaryentry{xmap}{name=XMAP,description={name of a pipelined MAP decoder},first=XMAP,firstplural=XMAPs}
\newacronym{xor}{XOR}{exclusive OR function}


\newacronym{zf}{ZF}{zero-forcing}

%% file: tex/abstract_neu_Norbert.tex
\glsunset{fpga}

\begin{abstract}
\glspl{dnn} are capable of solving complex problems in domains related to embedded systems, such as image and natural language processing.
To efficiently implement \glspl{dnn} on a specific \gls{fpga} platform for a given cost criterion, e.g. energy efficiency, an enormous amount of design parameters has to be considered from the topology down to the final hardware implementation. 
Interdependencies between the different design layers have to be taken into account and explored efficiently, making it hardly possible to find optimized solutions manually. 
An automatic, holistic design approach can improve the quality of \gls{dnn} implementations on \gls{fpga} significantly.
To this end, we present a cross-layer design space exploration methodology. 
It comprises optimizations starting from a hardware-aware topology search for \glspl{dnn} down to the final optimized implementation for a given \gls{fpga} platform.
The methodology is implemented in our \gls{half} framework, which combines an evolutionary search algorithm, various optimization steps and a library of parametrizable hardware \gls{dnn} modules. 
\gls{half} automates both the exploration process and the implementation of optimized solutions on a target \gls{fpga} platform for various applications. We demonstrate the performance of HALF on a medical use case for arrhythmia detection for three different design goals, i.e. low-energy, low-power and high-throughput 
respectively. Our \gls{fpga} implementation outperforms a TensorRT optimized model on an Nvidia Jetson platform in both throughput and energy consumption.

\glsresetall

\end{abstract}

%% file: tex/instroduction_neu_Norbert.tex
\glsunset{fpga}
\vspace{-1.5ex}
\glsresetall
\section{Introduction}
\label{sec:intro}

\begin{figure}[!h]
\centering
\includegraphics[width=1\columnwidth]{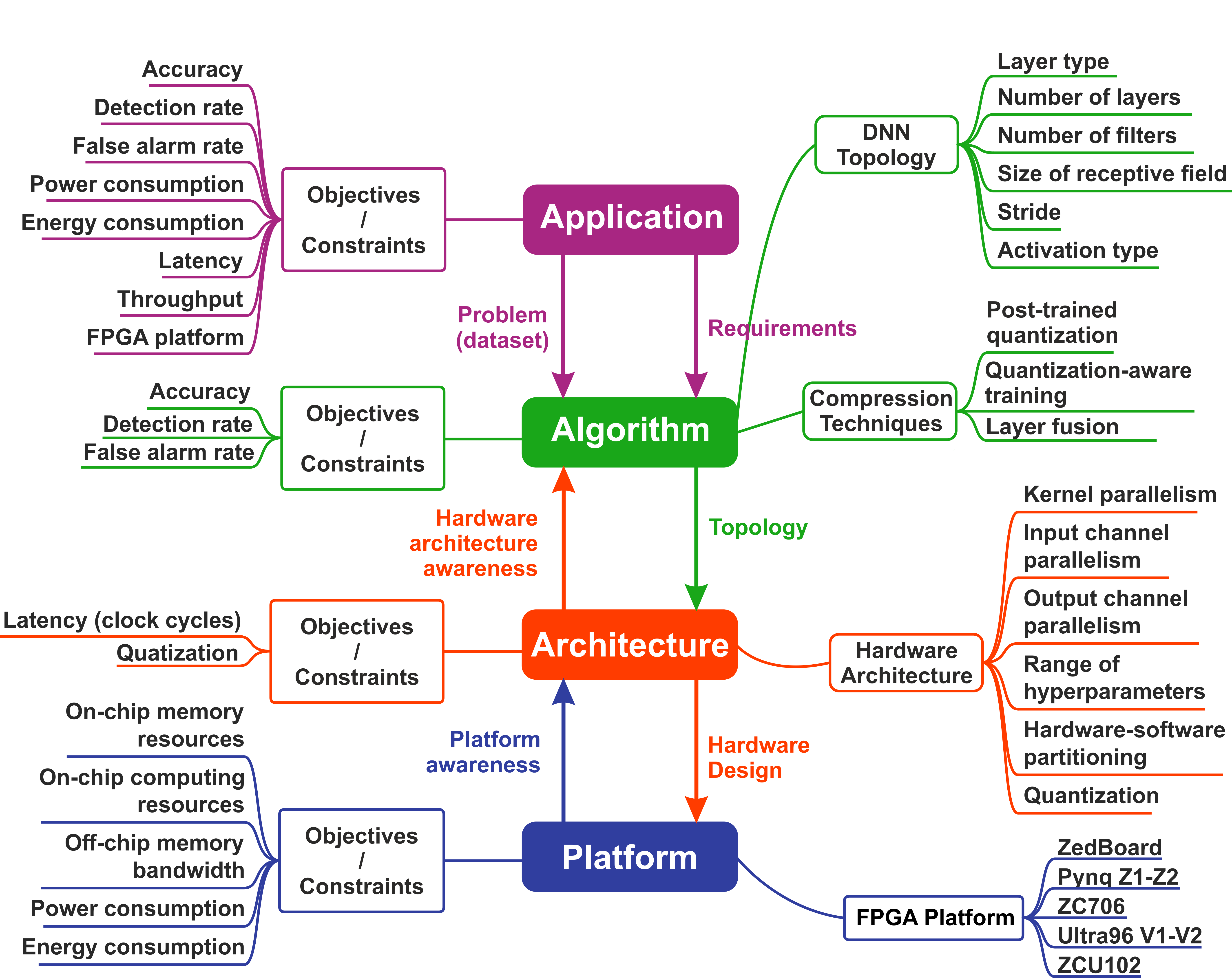}
\caption{Design hierarchy and multi-layer design space.}
\label{fig:design_space}
\end{figure}

Efficient implementation of \glspl{dnn} in hardware requires rigorous exploration of the design space on different layers of abstraction, including \textit{algorithmic}, \textit{architectural} and \textit{platform} layers, as it is depicted in \cref{fig:design_space}.
First, the \textit{application} is defined by a dataset and requirements in terms of \textit{constraints} and \textit{objectives}, e.g., accuracy, latency, etc.
At the highest level in the design hierarchy is the \textit{algorithm}, which is the most abstract description of the data and control flow in the form of a \gls{dnn} \textit{topology}.
The \textit{architecture} layer maps the topology to a \textit{hardware design}, which is implemented on the platform.
At the lowest level is the \textit{platform}, which describes the hardware and its physical properties.
All design layers introduce a large number of design choices as well as interdependencies (see \cref{fig:design_space}).

In our methodology, the architectural layer is represented by a highly parametrizable architecture template that is used to instantiate various \gls{dnn} topologies in hardware.
We formulate hardware models for latency, power, and energy derived from the architecture template expressed in terms of topology hyperparameters to include hardware awareness in the algorithmic layer.
The hardware models together with the architecture template describe a bridge between the algorithmic and platform layer.
However, this one bridge is not enough, since the dependence of the application to the architecture layer cannot be formalized.
To this end, we use \gls{nas} in the algorithmic layer and augment it with the hardware-awareness models of the architecture layer.
The \gls{nas} performs a full cross-layer optimization, which is guided by optimization objectives targeting both application requirements and hardware performance.
In summary, the hardware-aware \gls{nas} in strong coupling with the architecture template and its modeling spans a bridge from the application down to the platform layer, which allows a fully automatic design flow for optimized implementations.

The methodology is implemented in the \gls{half} framework, which comprises a hardware-aware evolutionary \gls{nas} and an \gls{fpga} implementation framework plus a hardware library.
We demonstrate the efficiency of our approach in a case study on energy-efficient \gls{fpga} implementations for atrial fibrillation detection in a real-world application scenario using a dataset provided by the Charité in Berlin. The novel contributions of the paper are:

\begin{itemize}	

\item A design space exploration methodology, which enables cross-layer optimization of \gls{dnn} for efficient hardware implementation.

\item A framework that automatically produces low-energy, low-power or high-throughput \gls{fpga} solutions.

\item A library of parameterizable low-power, ultra-low-latency hardware architectures of \gls{dnn} layers.

\item We demonstrate \gls{fpga} implementations for arrhythmia detection targeting different application scenarios which outperform an embedded GPU with respect to throughput, power and energy efficiency.

\end{itemize}

%% file: tex/related_works.tex
\section{Related work}
\label{sec:related_work}

\subsection{Neural Architecture Search}
\label{subsec:half_related_work_nas}

For the task of anomaly detection in \gls{ecg} data, \cite{kiranyaz2015real, rajpurkar2017cardiologist, hannun2019cardiologist, kachuee2018ecg} demonstrate that \glspl{1dcnn} can be used with raw \gls{ecg} samples without any hand-crafted feature extraction.
Instead of manually designing \gls{dnn} topologies for a specific detection task, \gls{nas} describes methods which automatically explore the search space spanned by \glspl{dnn} architectures.
Among the most successful techniques are gradient-based and evolutionary methods.
Recently, there is less activity around \gls{rl} based approaches, as both gradient-based and evolutionary methods can produce similar results in often significantly less time \cite{wu2019fbnet, real2019regularized}.
Nevertheless, \gls{rl} methods are explored in combination with hardware-awareness for hardware accelerators \cite{zhou2021rethinking} and \glspl{fpga} \cite{jiang2020hardware}.
The most prominent gradient-based method is \gls{darts} \cite{liu2018darts}, which searches for subgraphs in a larger, differentiable supergraph.
\cite{wu2019fbnet} introduced hardware-awareness to \gls{darts} in the form of latency minimization for mobile phones.
The authors of \cite{fan2020optimizing} used a very similar \gls{darts} setup as \cite{wu2019fbnet}, but for \glspl{fpga}, and the layer latencies were modelled as functions of the topology hyperparameters instead of lookup tables.
Although \gls{darts} is very fast, the structure of the manually designed supergraph can impose a strong bias.
In contrast, evolutionary algorithms do not require a supergraph.
Genetic algorithms use an encoding to describe the structure of the \gls{dnn}s, which enables biologie inspired concepts like crossover \cite{sun2020automatically} and aging \cite{real2019regularized} to be implemented.
Network morphisms are a class of operators on the graph structure, which change the \gls{dnn} such that retraining from scratch is not necessary \cite{wistuba2018deep, elsken2018efficient}. 
\cite{elsken2018efficient} uses network morphisms in their LEMONADE algorithm, which uses a bayesian method to generate \glspl{dnn} in a multi-objective search, although not hardware-aware.
Both \cite{elsken2018efficient} and \cite{jiang2020hardware} distinguish computationally "cheap" and "expensive" objectives, and skip unnecessary, "expensive" computations of bad candidates.
Later, \cite{schorn2020automated} augmented LEMONADE with hardware-awareness and error resilience.

\subsection{Automatic Hardware Generation and Hardware Architectures for \gls{dnn}}
\label{subsec:half_related_work_hardware}

Among the most used frameworks for automatic hardware generation are FINN \cite{umuroglu2017finn}, Xilinx ML Suite \cite{xilinx_ml_suite_overview} that is a toolchain for development on xDNN \cite{xdnn} general processing engine, and Deep Neural Network Development Kit (DNNDK) \cite{dnndk} that is an SDK for Deep Learning Processor Unit (DPU) \cite{dpu} programmable engine. However, none of them is a fully automatic approach, but rather a collection of tools that help to convert a \gls{dnn} into a custom \gls{fpga} accelerator, like FINN or compile them to a sequence of instructions executed on a programmable engine. All frameworks provide tools for \gls{dnn} compression and optimization and runtime support.

Both Xilinx ML Suite and DNNDK target \gls{dnn} execution on programmable engines that are designed to support a wide range of \gls{dnn} topologies. They trade off flexibility for generality. Contrary, FINN uses an \gls{hls} hardware library \cite{finn_hlslib} of hardware layers and components that are used to generate streaming architectures customized for each network. Other tools for automatic hardware generation are FlexCNN \cite{FlexCNN}, integrating an \gls{fpga} implementation framework into Tensorflow and DNNBuilder \cite{DNNBuilder}, which uses software-hardware co-design to perform an end-to-end optimization of deep learning applications.

Our approach is similar to FINN. Specifically, it maps \gls{dnn} on a set of highly optimized hardware components. Conceptually, the \gls{half} framework is different from all previous approaches as it includes \gls{nas} for hardware-optimized topologies. Also, we are targeting a fully automatic solution.

There are publications on augmenting \gls{nas} with hardware awareness for \gls{fpga}. \cite{fan2020optimizing} proposed \gls{nas} for their accelerator capable to process \glspl{cnn} layer-by-layer similar to xDNN and DPU. With respect to hardware awareness, their approach optimizes \glspl{dnn} for low latency only. \cite{jiang2020hardware} proposed a hardware and software co-exploration framework that uses \gls{nas} for optimizing \glspl{dnn} for implementation on multiple \glspl{fpga}, however without much consideration of optimizations on a level of separate \glspl{fpga}. Their approach is primarily focused on optimizing for high throughput.

Our framework is different as we use hardware-aware \gls{nas} for dataflow-style fully on-chip architectures customized for each network and optimized for various objectives, namely low latency, low power, and low energy.

%% file: tex/framework_neu_Norbert.tex
\section{\gls{half} Framework}
\label{sec:framework}

\begin{figure}[!h]
\centering
\includegraphics[width=0.9\columnwidth]{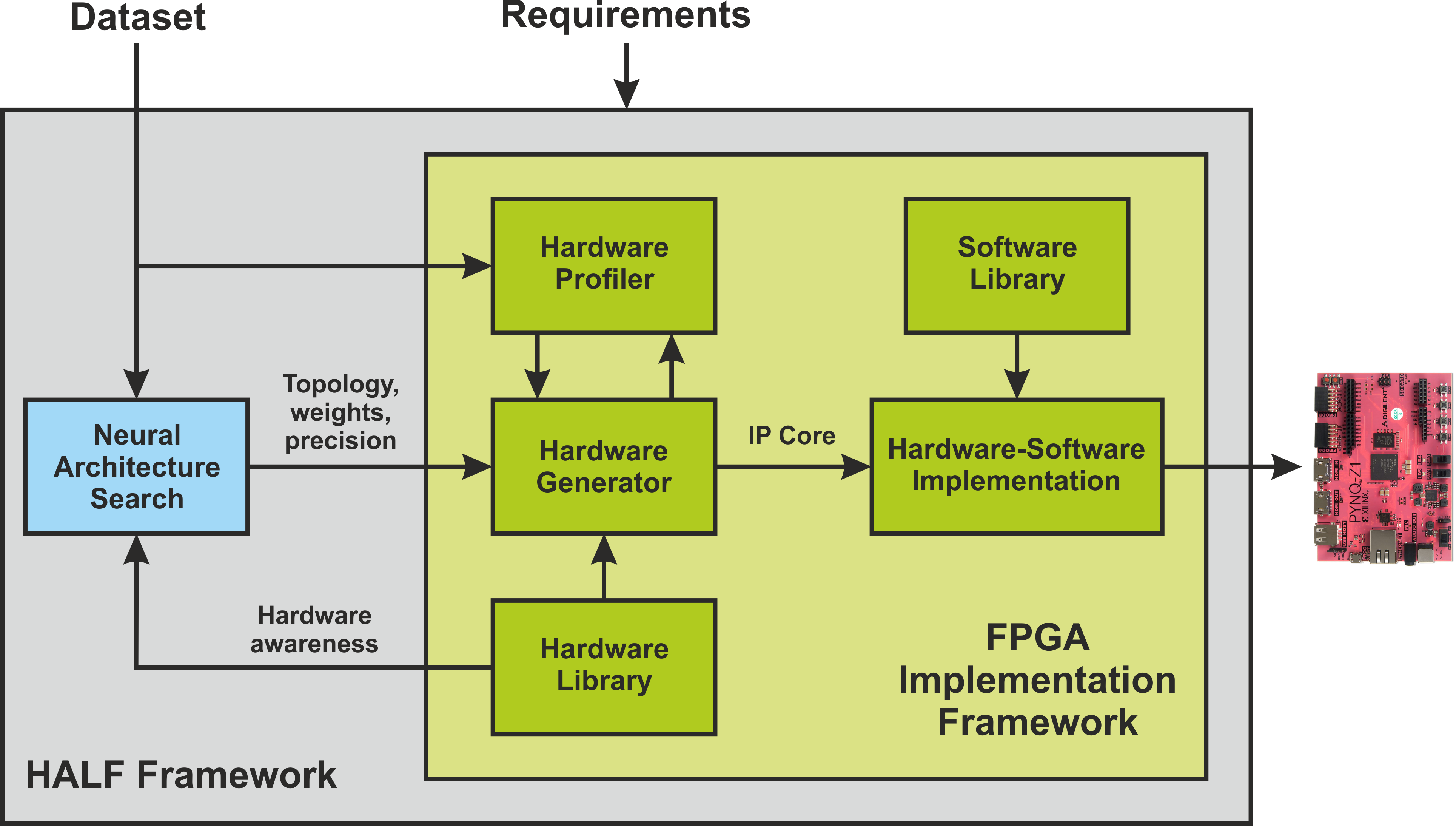}
\caption{Components of \gls{half} framework}
\label{fig:framework_details}
\end{figure}

The \gls{half} framework is comprised of two main components, which are the hardware-aware \gls{nas} and the \gls{fpga} implementation framework (see \cref{fig:framework_details}). 
The inputs are a dataset and requirements specified in terms of application-level and hardware-level constraints and optimization objectives. The output is a hardware configuration for the selected \gls{fpga} platform that fulfills the requirements.

\gls{half} generates the output automatically, and therefore it significantly accelerates the deployment of \glspl{dnn} on \glspl{fpga}. 
The \gls{nas} takes approximately two days, depending on the complexity of the underlying search space and the dataset, while manual search would take weeks, even without considering hardware awareness. 
Including hardware awareness into the \gls{nas} shortcuts the otherwise time-consuming manual design and evaluation cycles of different \gls{fpga} implementations to identify candidates with the best trade-offs. 
The automatic hardware generation and implementation only take a few hours, in contrast to a manual hardware design process that can take days or even weeks, especially if hardware components have to be designed from scratch.

\subsection{Neural Architecture Search}
\label{subsec:half_framework_nas}

The \gls{nas} is the first step in the framework and finds optimal topologies for the implementation framework.
It is based on an evolutionary algorithm, which are very flexible, as they do not impose strong restrictions on the search space or the objective functions, especially the latter do not need to be differentiable.
We use the genetic algorithm proposed by \cite{suganuma2017genetic}, which boosts the search via the concept of dormant genes.

For the selection strategy, we use a similar, bayesian-based method as \cite{elsken2018efficient}, which explores the Pareto Frontier of \gls{dnn} candidates efficiently in a two-step procedure, preselecting candidates based on computationally inexpensive objectives first.
Additionally to this two-step procedure, \cite{elsken2018efficient} uses network morphisms to increase the throughput of fully evaluated \glspl{dnn}.
We do not use network morphisms, which limits the range of mutation operations and also would be a bad fit for our genetic encoding. 
Instead, we handle the large training workload by implementing a dynamic workload scheduler, which leverages parallel processing on \gls{hpc} systems.

Hardware-awareness is incorporated twofold, i.e. via the search space and the optimization objectives.
The search space is constrained to layers which are included in the hardware library, thus the models from the \gls{nas} are guaranteed to be mappable to the device.
This encompasses aspects like layer types and valid hyperparameter combinations, but also the quantization of the inputs, weights and feature maps.
The second dimension of hardware-awareness is introduced by the optimization objectives, described in section \ref{sec:hardware_aware_objectives}.
Before passing the found topology with its trained weights to the hardware implementation framework, preprocessing and tuning techniques such as batchnorm-folding are applied to further compress the model.

\subsection{\gls{fpga} Implementation Framework}
\label{subsec:half_framework_fpga}

The \gls{fpga} implementation framework comprises a hardware generator, a custom hardware library, a profiler, a software library, and a hardware-software implementation step. The hardware generator produces a hardware architecture of the neural network using components from the hardware library described in \cref{sec:hardware_library}. Using Xilinx Vivado HLS, the framework generates an \gls{ip} core from the \gls{dnn} topology, which is based on the layers of the hardware library. The model weights are also integrated into the \gls{ip} core at this point because the hardware architecture uses only on-chip memory for model storage. Additionally, it instantiates interfaces for communication with external memory and \gls{fifo} buffers for connecting the elements. The hardware generator also calculates parallelization factors for each layer, which are based on the required throughput and are mainly constraint by the target platform, i.e., number of resources, available memory bandwidth, \gls{fpga} model. While the quantization of weights and activations is provided by the \gls{nas}, the quantization for the internal accumulators is found by profiling. The profiler identifies the optimal range and precision for all accumulators in the hardware and sets the bit widths accordingly. In the last hardware-software implementation step, Xilinx Vivado Design Suite is used to generate a bitstream for \gls{fpga}. The software is compiled for running on the processor cores of the board that is used to transfer input and output data to the \gls{fpga} and to control the \gls{ip} core.

%% file: tex/hardware_aware_objectives.tex
\section{Hardware-aware Objective Functions}
\label{sec:hardware_aware_objectives}

We choose energy, power and latency as the hardware-aware objectives and model them as functions of the topology and \gls{fpga} parameters.
The latency is defined as
\begin{equation}
\label{eq:t_total}
t_\text{total} = \sum_{j=1}^{N}\left(n_{\text{in},j}-1\right) \sigma_{j-1} + l_j,
\end{equation} 
where $N$ is the number of layers, $n_{in,j}$ is the number of values to initially fill the input buffer (e.g. the kernel size in case of convolution layers), $\sigma_{j-1}$ is the output rate of the previous layer in clock cycles and $l_j$ is the latency of the layer to produce its output.
Notice that $\sigma_{i} = \max(l_i, \sigma_{i-1})$ evaluates recursively and describes the pipelined nature of the hardware architecture.
The latencies $l_i$ depend on the layer type and hyperparameters such as strides and kernel size, but also loop unrolling factors $\alpha_i$ of the \gls{fpga} implementation.
In section \ref{sec:results}, the results are reported using the throughput instead of latency, which includes the contribution of data parallelism and is the batch size divided by the latency $t_\textrm{total}$.

We model the effective power, so that the energy can be simply described as the product of the runtime and the effective power.
The total power consumption is
\begin{equation}
\label{eq:total_power_comp}
P_\text{total} = P_\text{mem} + P_\text{board} + P_\text{stat} + P_\text{dyn}.
\end{equation} 
$P_\textrm{mem}$ is the power from memory transactions and mostly does not depend on the topology, because all the weights and activations are kept on the chip in our architecture. The size of the input sample, however, does influence $P_\textrm{mem}$, but since an input must always be read and cannot be optimized, its contribution to the power model is excluded. 
A model for $P_\textrm{mem}$ can be added easily to the framework, though.
$P_\textrm{board}$ is from other peripheral components of the hardware platform, but since it cannot be influenced by topology parameters, it is not modeled.
$P_\textrm{stat}$ and $P_\textrm{dyn}$ are the static and dynamic power consumption of the architecture, which we model in the total power with
\begin{equation}\label{eq:total_power}
P_\text{total} = \sum_{i=1}^{N} \alpha_i P_{\text{idle},i}^* 
    + \alpha_i \frac{t_{a,i}}{t_{\text{total}}}  P_{\text{calc},i}^*.
\end{equation}
We assume that the power scales linearly with the loop unrolling factors $\alpha_i$.
$P_{\textrm{idle},i}^*$ and $P_{\textrm{calc},i}^*$ are the power consumption when the layer is idling and calculating, respectively, for an unrolling factor of one, which can be estimated from the hardware profiler of the \gls{fpga} implementation framework.
$t_{a,i}$ is the time a layer is actively computing and it is the product of the total number of outputs the layer produces multiplied with the latency to produce one such output. 
The power can be minimized by using no unrolling (min $\alpha$) and stretching out the total runtime, with compliance to latency constraints.

The total energy is the product of the effective total power with the total runtime
\begin{equation}\label{eq:total_energy}
E_\text{total} = t_\text{total} P_\text{total} =  \sum_{i=1}^{N} \alpha_i t_\text{total} P_{\text{idle},i}^* 
    + \alpha_i t_{a,i} P_{\text{calc},i}^*.
\end{equation}
Looking at \cref{eq:total_energy}, it appears that minimizing the $\alpha_i$ is the best strategy to minimize the energy. However, since high $\alpha_i$ reduce both $t_\textrm{total}$ and $t_{a,i}$ superlinearly, it is high unrolling factors that reduce the total energy consumption.
Also, the energy consumed by the entire platform is the product of $P_\textrm{board}$ and $t_\textrm{total}$. 
Since $P_\textrm{board}$ can be much larger than the other contributions to $P_\textrm{total}$, minimizing $t_\textrm{total}$ is the most effective way to reduce the measurable energy consumption.

%% file: tex/hardware_library.tex
\section{Hardware Library}
\label{sec:hardware_library}

\begin{figure}[!h]
\centering
\includegraphics[width=0.9\columnwidth]{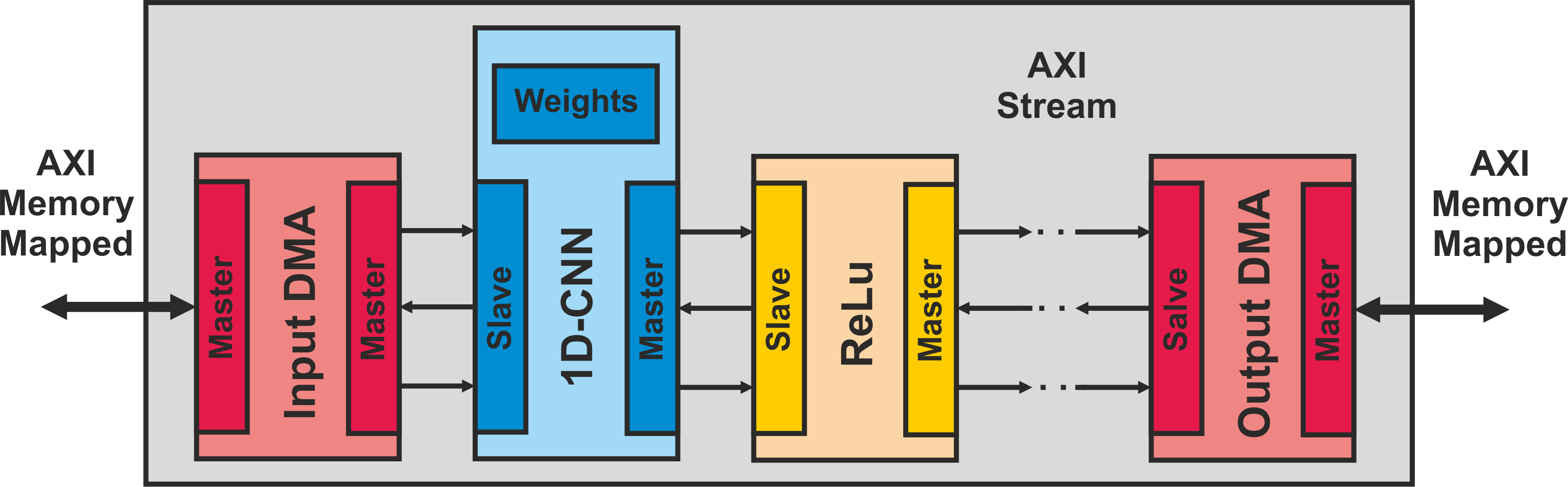}
\caption{Deep Neural Network Unit (DNNU).}
\label{fig:dnnu_arcitecture}
\end{figure}

We present an \gls{hls} hardware library of custom hardware architectures for standard \glspl{1dcnn}, depth-wise separable \glspl{1dcnn} and various other \gls{dnn} layers and components. The hardware architectures are highly customizable, which allows the implementation of various neural topologies. The hardware library is written as a collection of C++ template functions with \gls{hls} annotations and modularity in mind to make it easily expandable by new layers. The hardware architecture is designed to be low power and ultra-low latency. Primarily, this is achieved by keeping all weights and intermediate results in on-chip memory since off-chip transfers consume more energy and introduce extra latency. External memory is only used to read input data and write results, reducing memory access to the absolute minimum. The architecture is fully pipelined, allowing all layers to operate concurrently and starting the computation as soon as the inputs are ready to reduce latency and energy consumption. The library is based on dataflow architectures, which can be easily customized for each network. The hardware modules are designed with streaming interfaces to facilitate fast design, debugging, interoperability, and ease of integration. Separate hardware modules dedicated to each layer are connected using on-chip data streams in a single top-level module called \gls{dnnu}, as shown in \cref{fig:dnnu_arcitecture}. The top-level module is equipped with \gls{dma} components that allow access to external memory independent of any processor using AXI-Master interfaces.

In a pipelined architecture, there always exists a bottleneck stage, which determines the latency of the entire pipeline. The latency of the bottleneck stage can be decreased by spatial parallelism, which we refer to as unrolling (coming from loop unrolling). The hardware library is designed with parametrizable unrolling, which parallelizes the bottleneck stages efficiently. The parametrization allows coarse-grained parallelization on a level of filters for \gls{cnn} layers and neurons for fully connected layers, and fine-grained parallelization on a level of dot-products, distinguishing kernel-level and input-channel parallelism.

%% file: tex/results_neu_Norbert.tex

\section{Results}
\label{sec:results}

The effectiveness of our methodology is evaluated on the task of a binary classification for \gls{ecg}-based arrhythmia detection. The dataset, provided by the university hospital Charité in Berlin,  contains 16000 samples, with 2 channels and a length of 60000 each. The dataset contains equal amounts of positive and negative samples.
The task performance is measured in detection and false alarm rate, where we define hard limits of 90\,\% and 20\,\% for acceptance, respectively.

The \gls{nas} is performed on four Nvidia Titan X GPUs for 100 generations with 20 children per generation, which takes two days to finish.
The search space constitutes of depth-wise separable convolutions with 60 different hyperparameter configurations and max pooling with 4 different strides.
All \glspl{dnn} end with a global average-pooling layer followed by a fully-connected layer.
The depth of the topology is chosen by the \gls{nas} but restricted between 2 and 15 layers (final layers not included).
The optimization objectives are power, energy and latency each with and without unrolling, and additionally number of parameters, detection and false alarm rate. All objectives are considered at the same time in the Pareto frontier.

\subsection{Influence of \gls{nas} objectives on the topology}

The network topology itself has a high impact on the final performance in terms of power and energy, which we show by comparing models optimized for the three different objectives of low-power with minimal parallelization (low P, min $\alpha_{\scaleto{\text{NAS}}{3.5pt}}$), low-energy with minimal parallelization (low E, min $\alpha_{\scaleto{\text{NAS}}{3.5pt}}$) and low-energy with maximal parallelization (low E, max $\alpha_{\scaleto{\text{NAS}}{3.5pt}}$).
The min $\alpha_{\scaleto{\text{NAS}}{3.5pt}}$ case is applied whenever the hardware resources are relatively low compared to the size of the \gls{dnn} model, so full unrolling is not possible.
The case (low P, max $\alpha_{\scaleto{\text{NAS}}{3.5pt}}$) is excluded, since the objective of low power and maximum unrolling leads to unreasonable topologies, which can hardly be unrolled by design.
The models optimized for high-throughput are the same as the ones for energy, thus not explicitly considered here.
For the implementation strategy, we set the parallelization factor $\alpha_{\scaleto{\text{Impl.}}{5pt}}$ to either one or the maximum, while keeping other hardware-related parameters fixed.

\begin{table}[!h]
\centering
\caption{Power and energy measurements using different objectives and resulting topologies.}
\label{tab:topology_results}
\begin{tabular}{ccccc}
\toprule
\begin{tabular}[c]{@{}c@{}}\gls{nas}\\Objective\end{tabular} &
\begin{tabular}[c]{@{}c@{}}Impl.\\Strategy\end{tabular} &
\begin{tabular}[c]{@{}c@{}}Throughput\\ {[}samples/s{]}\end{tabular} &
\begin{tabular}[c]{@{}c@{}}$P_\text{total}$\\ {[}W{]}\end{tabular} &
\begin{tabular}[c]{@{}c@{}}$E_\text{total}$\\ {[}\textmu J{]}\end{tabular} 
\\ 
\midrule

low E, max $\alpha_{\scaleto{\text{NAS}}{3.5pt}}$ & min $\alpha_{\scaleto{\text{Impl.}}{5pt}}$ & 4.4$\cdot 10^{3}$ & 4.42 & 1010  \\ 

low E, min $\alpha_{\scaleto{\text{NAS}}{3.5pt}}$ & min $\alpha_{\scaleto{\text{Impl.}}{5pt}}$ &  \textbf{5.3}$\mathbf{\cdot 10^{3}}$ & 4.46 & \textbf{841} \\

low P, min $\alpha_{\scaleto{\text{NAS}}{3.5pt}}$ & min $\alpha_{\scaleto{\text{Impl.}}{5pt}}$ & 1.4$\cdot 10^{3}$ & \textbf{4.40} & 3120 \\

\midrule

low E, max $\alpha_{\scaleto{\text{NAS}}{3.5pt}}$ & max $\alpha_{\scaleto{\text{Impl.}}{5pt}}$ & \textbf{4.8}$\mathbf{\cdot 10^{5}}$ & 8.22 & \textbf{1.7} \\

low E, min $\alpha_{\scaleto{\text{NAS}}{3.5pt}}$ & max $\alpha_{\scaleto{\text{Impl.}}{5pt}}$ & 3.7$\cdot 10^{5}$ & 7.16 & 2.0 \\

low P, min $\alpha_{\scaleto{\text{NAS}}{3.5pt}}$ & max $\alpha_{\scaleto{\text{Impl.}}{5pt}}$ & 8.3$\cdot 10^{4}$ & \textbf{6.10} & 73.8 \\

\bottomrule
\end{tabular}
\end{table}

\cref{tab:topology_results} shows that the best results in terms of energy and power are obtained if the optimization objective matches the implementation strategy. 
This demonstrates the effectiveness of the cross-layer optimization approach, where hardware-related parallelization factors influence the topology search, leading to better solutions. 
\begin{figure}[]
\centering
\includegraphics[width=0.9\columnwidth]{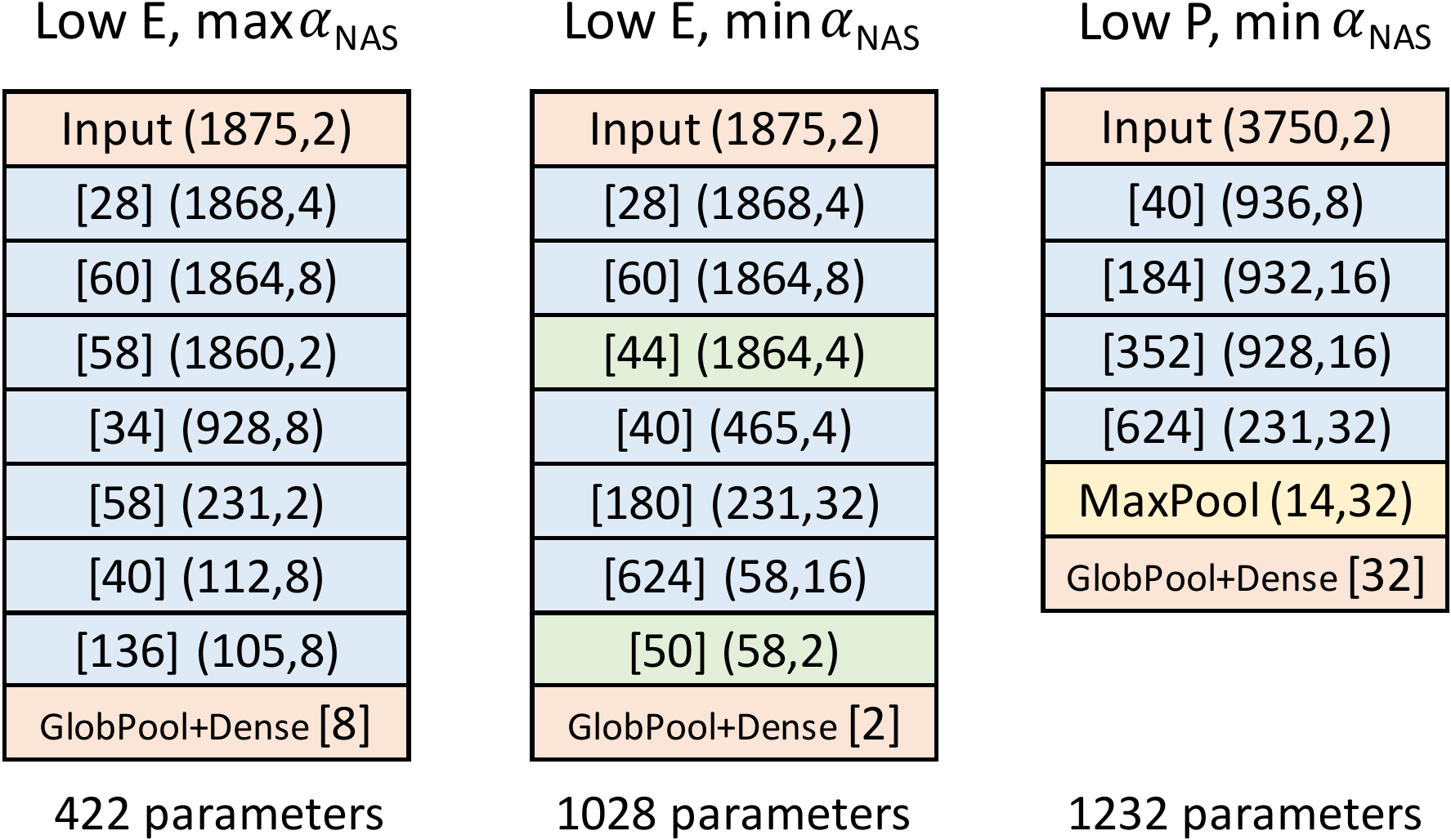}
\caption{Comparison of topologies for different optimization objectives. Blue: depthwise-separable convolution, green: kernel size 1. In the boxes: [\#parameters], (output-width, channels)}
\label{fig:topo_comparison}
\end{figure}
The three topologies used in \cref{tab:topology_results} have meaningful differences in their structure, as shown in \cref{fig:topo_comparison}.
For the first two models, both optimized for energy, the search converged to a very similar solution.
Especially the first layers are identical and the position of the striding layers is the same.
A parent-child relation in terms of evolution is ruled out since we selected both models from two different experiments.
The key difference comes from the \nth{3} convolution layer, which becomes the bottleneck in the min $\alpha_{\scaleto{\text{Impl.}}{5pt}}$ case, which it is not for max $\alpha_{\scaleto{\text{Impl.}}{5pt}}$.
The \nth{2} model uses a kernel size of one here, which lowers the latency of the entire pipeline significantly, resulting in higher energy-efficiency, although having \textit{two times} more parameters.
For the case of max $\alpha_{\scaleto{\text{Impl.}}{5pt}}$ the \nth{1} model has lower energy consumption, because it needs less resources, thus more parallel instances can be implemented and the throughput is increased.
The \nth{3} topology in \cref{fig:topo_comparison} is less deep than the energy optimized models.
Without unrolling, a shallower topology requires less hardware resources.
The bottleneck layer is the \nth{3} convolution, which computes twice as long as the next slowest layer.
This imposes a high idle time on the non-bottleneck layers, which results in lower power consumption, although having almost \textit{three times} as many parameters as the \nth{1} model.

In summary, it is not the size of the model alone, but its structure, which determines the hardware performance.
The \gls{nas} is able to find significant structural features in the \gls{dnn} models, based on the optimization objectives.

\subsection{Efficiency of holistic methodology}

To demonstrate the efficiency of our holistic approach, we present solutions for three different domains, namely low-power, low-energy, and high-throughput with optimizations applied on all design levels from the \gls{nas} down to the \gls{fpga} platform, see \cref{tab:final_solutions}. In each case, the target platform was selected according to the optimization goal, Pynq-Z1, Ultra96-V2, and ZCU102 for each domain, respectively. Additionally, we show results for the low-energy topology implemented on the Nvidia Jetson AGX Xavier embedded GPU optimized using TensorRT. 

For the low-power domain, the \gls{nas} searched for a topology that exhibits the lowest power with unrolling factor constrained to one (low P, min $\alpha$). In its turn, the hardware implementation framework instantiated only a single fully-folded instance of \gls{dnnu} implemented with the lowest frequency that, however, still outperforms real-time requirements. Although the \gls{nas} includes separate objectives for low-energy and high-throughput, we observe that the best model for both cases is the same one, which is optimized considering the maximal unrolling factor (low E, low L, max $\alpha$). The following hardware implementation step targeted different platforms but used identical strategies to achieve the highest frequency and maximally utilize the available resources by instantiating the maximal number of instances with the highest unrolling factor ($\alpha$ = 40). \cref{tab:final_solutions} demonstrates that each implementation achieves the best results in the targeted domain. The \gls{fpga} designs outperform the embedded GPU implementation regarding all shown metrics, although the GPU has a higher frequency, larger batch size, and a model optimized with TensorRT.

\begin{table}[!h]
\centering
\caption{Comparison of \gls{fpga} implementations for various domains.}
\label{tab:final_solutions}
\begin{tabular}{@{\hspace{0.05cm}}l@{\hspace{0.1cm}}c@{\hspace{0.3cm}}c@{\hspace{0.3cm}}c@{\hspace{0.3cm}}c@{\hspace{0.3cm}}cc@{\hspace{0.05cm}}}
\toprule
Device &

\begin{tabular}[c]{@{}c@{}}Freq.\\ {[}MHz{]}\end{tabular} &
\begin{tabular}[c]{@{}c@{}}Batch\\ Size\end{tabular} &
\begin{tabular}[c]{@{}c@{}}Throughput\\ {[}samples/s{]}\end{tabular} &
\begin{tabular}[c]{@{}c@{}}$P_\text{total}$\\ {[}W{]}\end{tabular} &
\begin{tabular}[c]{@{}c@{}}$E_\text{total}$\\ {[}J{]}\end{tabular} \\

\midrule 
Pynq-Z1                                     &
0.5                                         &
1                                           &
2.1                                         &
\textbf{1.9}                                &
9.1$\cdot 10^{-1}$                          \\

\midrule 
Ultra96-V2                                  &
333                                         &
4                                           &
4.8$\cdot 10^{5}$                           &
8.22                                        &
\textbf{1.7}$\mathbf{\cdot 10^{-5}}$        \\

\midrule 
ZCU102                                      & 
322                                         &
16                                          &
\textbf{1.6}$\mathbf{\cdot 10^{6}}$         &
33.9                                        &
2.1$\cdot 10^{-5}$                          \\

\midrule 
Jetson AGX                                  &
1377                                        &
1024                                        &
7.7$\cdot 10^{4}$                           &
21.1                                        &
2.7$\cdot 10^{-4}$                          \\

\bottomrule
\end{tabular}
\end{table}

%% file: tex/conclusion_neu_Norbert.tex
\section{Conclusion}
\label{sec:conclusion}

We present a cross-layer optimization methodology, which allows searching for \glspl{dnn} optimized for hardware. The methodology is based on a hardware-aware \gls{nas} coupled with a parametrizable hardware template. 
We implemented this approach in an automatic framework and demonstrate its performance by comparing power and energy for \gls{dnn} models optimized for different objectives.
The objectives affect the structure of the generated networks meaningfully, so that the model implementations outperform each other in their respective optimization target. 
Additionally, we exploit the full potential of our framework by automatically applying hardware-related optimizations that further tune the model, depending on the target platform.
Considering every design aspect of the hardware implementation, we target different domain and show significant differences in the hardware metrics for a real-world application scenario of atrial fibrillation detection, outperforming the  Nvidia Jetson AGX in throughput, power and energy consumption.